\documentclass[letter]{jpsj3}
\usepackage{txfonts}
\title{Solitons in the Crossover between Band Insulator and Mott Insulator: 
Application to TTF-Chloranil under Pressure}

\author{Hidetoshi Fukuyama$^1$, and Masao Ogata$^2$}
\inst{
$^1$Research Institute for Science and Technology, Tokyo University of Science, 
Shinjuku, Tokyo 162-8601, Japan \\
$^2$Department of Physics, University of Tokyo,
Hongo 7-3-1, Bunkyo-ku, Tokyo 113-0033, Japan} 

\abst{
Based on the Phase Hamiltonian, two types of solitons are found to
 exist in the crossover region between band insulator and Mott insulator in one-dimension. 
Both of these solitons have fractional charges but with different spins, zero and 1/2, respectively. 
The results are in accord with the experimental results by Kanoda {\it et al.} for 
TTF-Chloranil under pressure.}

\begin{document}
\maketitle

Recent experiment by Kanoda {\it et al.} on TTF-Chloranil (TTF-QCl)\cite{Kanoda}
has revealed an intriguing feature of the phase diagram on the plane of 
pressure and temperature. 
Above all, at room temperature the non-magnetic neutral state at 
ambient pressure crossovers smoothly to magnetic ionic state without dimerization  
as a function of pressure at around $P_{\rm c}\sim 8$ GPa, where
both the conductivity along chains and the NMR relaxation rate 
show sharp changes.\cite{Kanoda} 
The experiment indicates that there is no lattice dimerization 
in the region of high pressure. 
Hence this experiment discloses the fact that there exists a smooth 
change from band insulator (BI) at low pressure to Mott insulator (MI) 
at high pressure with some characteristic changes of charge and spin 
excitations in the crossover region.


Defining $\varepsilon_{\rm c}$ and $\varepsilon_{\rm s}$ 
by conductivity$= {\rm exp}(-\varepsilon_{\rm c})$ S/cm and 
$1/T_1={\rm exp}(-\varepsilon_{\rm s})$ /sec, respectively, 
from the experiment,\cite{Kanoda} 
we see that $\varepsilon_{\rm c}$ first 
decreases as the pressure increases, has a minimum near $p=8$ kbar, 
and then increases saturating near $p=16$ kbar, while 
$\varepsilon_{\rm s}$ decreases as the pressure increases monotonically and 
gradually saturates above $p=10$ kbar. 
Moreover the temperature $(T)$ dependences of  both $\varepsilon_{\rm c}$ 
and $\varepsilon_{\rm s}$ are of activation type, i.e., 
proportional to inverse temperatures at least in the crossover region. 
Hence 
we expect that $E_{\rm c}=T\varepsilon_{\rm c}$ and 
$E_{\rm s}=T\varepsilon_{\rm s}$ at fixed temperature (room temperature) 
disclose essentially the pressure dependences of activation energy 
of charge and spin excitations, 
which we will explore in the following. 
We will employ the model of one-dimensional electrons, which will be 
justified by the experimental observation of strong anisotropic conductivity 
at least around the crossover pressure region. 


Our model is standard one\cite{RiceMele,Nagaosa,Fabrizio,Tsuchi} with on-site $U$, 
and inter-site, $V$, Coulomb repulsion together with alternating site potentials, $\Delta$, 
reflecting TTF and QCL sites ($\Delta$ is assumed to be positive). 
The average electron number per site is one, i.e., half-filled band.
\begin{equation}
\begin{split}
{\cal H} &= -\sum_{\ell,\sigma} t_{\ell,\ell+1} \left( 
c_{\ell\sigma}^\dagger c_{\ell+1,\sigma} + c_{\ell+1,\sigma}^\dagger c_{\ell\sigma} \right) \cr
&+ \frac{\Delta}{2} \sum_{\ell} (-1)^\ell n_{\ell} + U\sum_{\ell} n_{\ell\uparrow} n_{\ell\downarrow} 
+ V \sum_{\ell} n_{\ell} n_{\ell+1}. 
\label{Hamilt}
\end{split}
\end{equation}
Here $c_{\ell\sigma} (c_{\ell\sigma}^\dagger)$ is the electron annihilation (creation) 
operator, $n_{\ell\sigma}=c_{\ell\sigma}^\dagger c_{\ell\sigma}$ with spin $\sigma$, 
the electron number operator at the site $\ell$,
and $n_{\ell} = n_{\ell\uparrow}+n_{\ell\downarrow}$. 
The physics of such interacting one-dimensional electrons can be treated 
in a transparent way by the Phase Hamiltonian\cite{Suzumura,HF} 
based on the bosonization first introduced by Tomonaga.\cite{Tomonaga} 
Equation (\ref{Hamilt}) can be transformed into the following by keeping 
the minimal interactions relevant to the stabilization of BI and MI.\cite{Tsuchi}
\begin{equation}
\begin{split}
{\cal H} &= \frac{v}{2\pi} \int dx \biggl[ 
\frac{1}{2} (\partial_x \theta)^2 + \frac{c}{2} (\partial_x \phi)^2
-\gamma_\Delta \sin\theta \cos\phi - \gamma_c \cos2\theta \biggr] \cr
&\equiv \frac{v}{2\pi} \int dx \biggl[ 
\frac{1}{2} (\partial_x \theta)^2 + \frac{c}{2} (\partial_x \phi)^2 +V(\theta,\phi) \biggr]. 
\label{Ene}
\end{split}
\end{equation}
where $\theta\ (\phi)$ are phase variables for charge (spin), 
$v=2ta+(U+6V)a/2\pi$ is the charge velocity, $c$ is the ratio $v_s/v$ 
with $v_s=2ta-(U-2V)a/2\pi$ being the spin velocity, and 
$\gamma_c = g_c/\pi a^2 v, \gamma_\Delta= g_\Delta/\pi a^2 v$ with
$g_c = (U-2V)a, g_\Delta=4\pi\Delta a$ where $a$ is the lattice constant. 
By noting that the system of present interest is at room temperature and 
then direct quantum effects will not be essential, we treat the phase Hamiltonian
classically keeping possible renormalization of $g_c$ and $g_\Delta$ in mind.

The states of local minimum energy in the space of $\theta$ and $\phi$ are determined 
by taking variation of the potential energy, $V(\theta,\phi)$, in eq.~(\ref{Ene}) 
relative to these phase variables resulting in 
\begin{align}
\frac{\partial V}{\partial \theta} 
&=- \gamma_\Delta \cos\theta \cos \phi + 2\gamma_c \sin 2\theta \nonumber \\
&= -\gamma_\Delta \cos \theta \left[ \cos\phi -4\alpha \sin\theta \right] =0, 
\label{EqA} \\
\frac{\partial V}{\partial \phi} &= \gamma_\Delta \sin\theta \sin \phi = 0, 
\label{EqB}
\end{align}
where $\alpha= \gamma_c/ \gamma_\Delta$. 

\begin{figure}
\includegraphics[width=10cm]{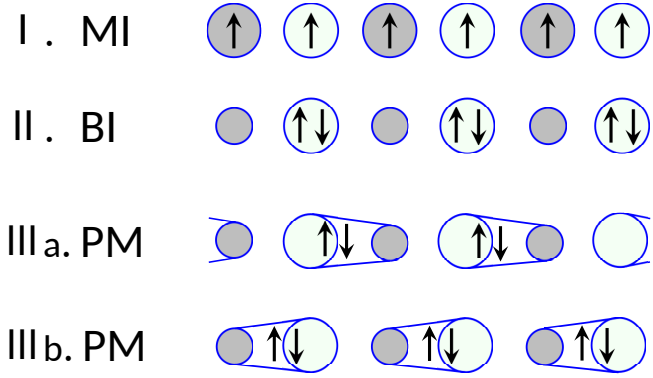}
\caption{Three kinds of solutions satisfying eqs.~(\ref{EqA}) and (\ref{EqB}). 
The state I corresponds to ideal MI and that the state II to BI. 
The states IIIa and IIIb correspond to the two kinds of 
polarized Mott (PM) Insulator (see the text).}
\label{Fig:1}
\end{figure}

\begin{figure}
\includegraphics[width=7cm]{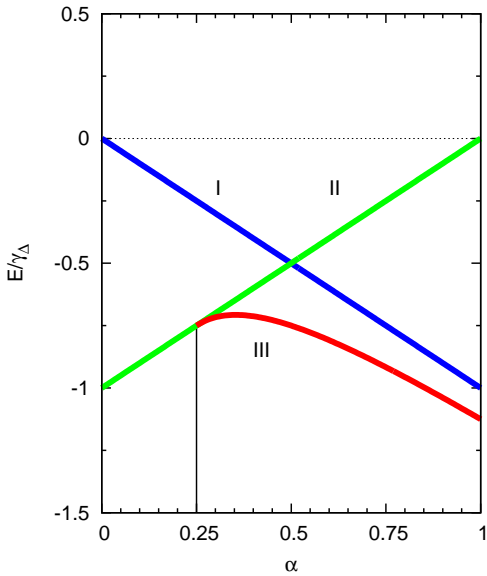}
\vskip 2cm
\caption{Energy of the three states as a function of 
$\alpha= \gamma_c/ \gamma_\Delta$ for a fixed $\gamma_\Delta$.}
\label{Fig:3}
\end{figure}

There exist three solutions satisfying eqs.~(\ref{EqA}) and (\ref{EqB}) 
simultaneously depending on the value of $\alpha$. 
For $\alpha<1/4$, 

I. $\cos\phi=0$, and $\sin\theta=0$, leading to $\phi = \pm \pi/2$; $\theta=0, \pi$,

II. $\sin\phi=0$, and $\cos\theta=0$, leading to $\phi=0,\pi$; $\theta = \pm \pi/2$, 
\par\noindent
and for $\alpha>1/4$, the third solution is possible, which is

III.	$\sin\phi=0$, and $\cos\phi - 4\alpha\sin\theta =0$, leading to 
$\phi=0$; $\sin\theta = 1/4\alpha$ (state IIIa) or 
$\phi=\pi$; $\sin\theta = -1/4\alpha$ (state IIIb). 
The characteristic features of these states are shown in Fig.~\ref{Fig:1}. 
It is seen that the state I corresponds to ideal MI and II to BI. 
The state III smoothly interpolates BI and MI, and is characterized by 
the coexistence of the charge density wave (CDW) and 
bond CDW\cite{Tsuchi} as shown in Fig.~\ref{Fig:1}
which we call polarized Mott (PM) insulator. 
The energy of these states (lower one among each kind) are
\begin{equation}
\begin{split}
V_{\rm I} &= -\gamma_c = - \alpha \gamma_\Delta \cr
V_{\rm II} &= \gamma_c - \gamma_\Delta = -(1-\alpha) \gamma_\Delta, \qquad 
({\rm for}\ \phi=0; \theta=\pi/2,  {\rm or}\ \phi=\pi; \theta=-\pi/2), \cr
V_{\rm III} &= - \left( \alpha + \frac{1}{8\alpha} \right) \gamma_\Delta. 
\label{EGR}
\end{split}
\end{equation}
The $\alpha$ dependence of the energy of these states is shown in Fig.~\ref{Fig:3}, 
which indicates the lowest energy state is II for $\alpha<1/4$, and III for $\alpha>1/4$. 
Confining to the region of $\alpha>1/4$, we will focus on the state III in the following.

\begin{figure}[h]
\includegraphics[clip,width=10cm]{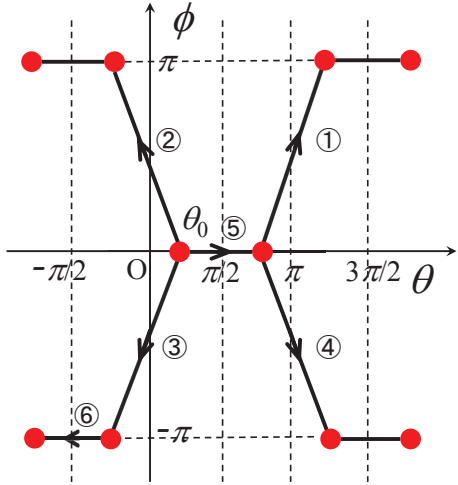}
\caption{Phase solitons between the states III in the plane of $\theta$ and $\phi$. 
The red solid circles indicate states with energy $V_{\rm III}$, which are degenerate. 
The arrows with number 1 to 6 indicate the six different and non-equivalent 
solitons; 1, 2, 3 and 4 are the solitons with spin $\pm 1/2$ and fractional charge, 
while 5 and 6 are spinless charge solitons with fractional charge.}
\label{Fig:4}
\end{figure}

The states III are indicated by red solid circles on the plane of $\theta$ and 
$\phi$ in Fig.~\ref{Fig:4}.  
All these states are degenerate. 
As seen, there exist four different and non-equivalent paths connecting 
these states with potential barriers in between, which are solitons. 

Soliton 1: spin $1/2$,  charge $2\theta_0/\pi$, 

Soliton 2: spin $1/2$, charge $-2\theta_0/\pi$, 

Soliton 3: spin $-1/2$,  charge $2\theta_0/\pi$, 

Soliton 4: spin $-1/2$, charge $-2\theta_0/\pi$, 

Soliton 5: spin $0$, charge $2(\pi/2-\theta_0)/\pi$,

Soliton 6: spin $0$, charge $-2(\pi/2-\theta_0)/\pi$,

\noindent
where $\theta_0=\sin^{-1} (1/4\alpha)$.

The solitons in the half-filled band had been studied in the presence of 
dimerizations,\cite{Hara,Naga2} but the present case without dimerization 
bridging directly between band insulator and Mott insulator had not been studied before.
The spatial form of these solitons are described by 
\begin{equation}
\begin{split}
&-\frac{\partial^2 \theta}{\partial^2 x} - \gamma_\Delta \cos\theta \cos \phi + 2\gamma_c \sin 2\theta =0. \cr 
&- c \frac{\partial^2 \phi}{\partial^2 x} + \gamma_\Delta \sin\theta \sin \phi = 0. 
\end{split}
\end{equation}
For solitons 5 and 6 with $\phi$ fixed at 0 or $\pm\pi$, 
the spatial dependence and formation energy of these solitons are given analytically as
\begin{equation}
\theta_5(x) = 2\tan^{-1} \left[ \sqrt{16\alpha^2-1} {\rm tanh} \left(
\frac{\sqrt{16\alpha^2-1}}{4\alpha} \sqrt{\gamma_c} (x-x_0) \right) +4\alpha \right] 
= -\theta_6(x), 
\end{equation}
and 
\begin{equation}
\begin{split}
E_5=E_6 
&= \frac{v}{2\pi} \frac{\sqrt{\gamma_c}}{\alpha} \left( \sqrt{16\alpha^2-1} - \frac{\pi}{2} + \theta_0\right) \cr
&= \frac{2v}{\pi} \sqrt{\gamma_c} \left( \cos\theta_0 - \sin \theta_0 \left( \frac{\pi}{2} - \theta_0\right) \right).
\end{split}
\end{equation}

\begin{figure}
\includegraphics[width=12cm]{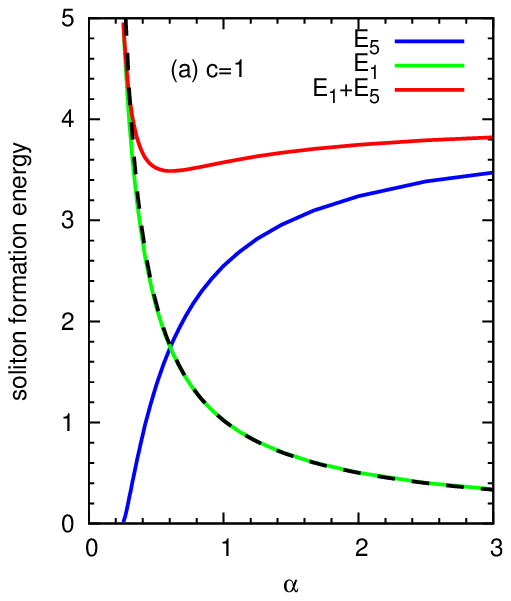}
\includegraphics[width=12cm]{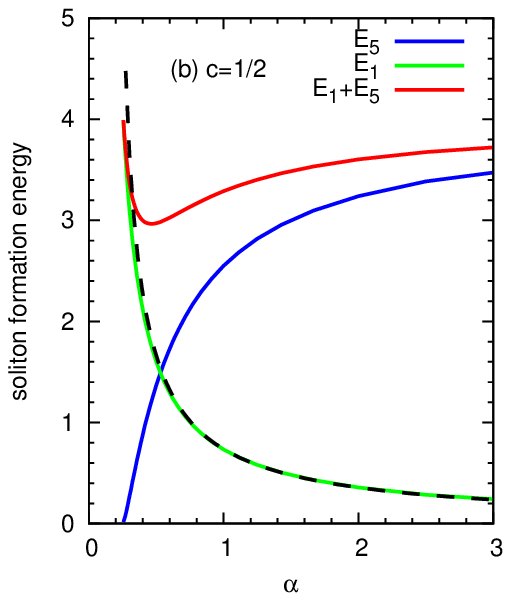}
\vskip 2cm
\caption{Formation energies of solitons as a function of 
$\alpha= \gamma_c/ \gamma_\Delta$ for the cases with (a) $c=1$ and (b) $c=1/2$. 
$E_1$ and $E_5$ indicate the formation energy for soliton 1 and 5 in Fig.~\ref{Fig:4}, respectively. 
Red curves show the sum of the two formation energies.
Broken lines for $E_1$ are the results of approximate solutions under the 
assumption of $\cos\phi=4\alpha\sin\theta$.}
\label{Fig:5}
\end{figure}

Those of solitons 1, 2, 3, and 4 can not be given analytically and the result of 
numerical calculation of formation energy, $E_1 (=E_2=E_3=E_4)$ 
are given in Fig.~\ref{Fig:5} as a function of $\alpha$ ($\alpha > 1/4$) 
together with that of soliton 5 (6). 
If the path of soliton 1 (2, 3, 4) are assumed to follow $\cos\phi = 4\alpha\sin\theta$, 
the spatial variations of 
$\theta_1(x)=\theta_1^{\rm A}$ ($\theta_2^{\rm A}, \theta_3^{\rm A}, \theta_4^{\rm A}$) 
and $\phi_1(x)=\phi_1^{\rm A}$ ($\phi_2^{\rm A}, \phi_3^{\rm A}, \phi_4^{\rm A}$) 
are given as follows together with the resultant formation energy, $E_1^{\rm A}$ 
($E_2^{\rm A}, E_3^{\rm A}, E_4^{\rm A}$), 
\begin{equation}
\begin{split}
\theta_1^{\rm A} (x) &= \sin^{-1} \left[ \frac{1}{4\alpha} \cos\phi \right] 
= \pi + \sin^{-1} \left[ \frac{1}{4\alpha} \tanh \left( \frac{\sqrt{\gamma_c} (x-x_0) }{2\alpha\sqrt{c}} \right) \right]
= \pi - \theta_2^{\rm A} (x) = \pi - \theta_3^{\rm A} (x) = \theta_4^{\rm A} (x), 
\cr
\phi_1^{\rm A} (x) &=2\tan^{-1} \left[ \exp \left( \frac{\sqrt{\gamma_c} (x-x_0) }{2\alpha\sqrt{c}} \right) \right]
= \phi_2^{\rm A} (x) = -\phi_3^{\rm A} (x)  = -\phi_4^{\rm A} (x) , \cr
E_1^{\rm A} &= \frac{v}{2\pi} \frac{\sqrt{\gamma_c}}{\sqrt{c}} 
\left[ \frac{c}{\alpha} + \frac{1}{2\alpha} -\left( 1-\frac{1}{16\alpha^2}\right) \log \frac{4\alpha+1}{4\alpha-1}
\right] =E_2^{\rm A} = E_3^{\rm A} =E_4^{\rm A}, 
\end{split}
\end{equation}
the last of which is shown in Fig.~\ref{Fig:5}  by dashed line for comparison.
It is seen that $E_1^{\rm A}$ is a good approximation for $E_1$. 
Note that $c$ is the ratio of spin velocity and charge velocity, $c=v_s/v$, and then
$c$ is smaller than 1 for the case of repulsive interactions. 
Although $c$ can be dependent on the pressure, we have shown typical cases with 
$c=1$ and $c=1/2$ in Fig.~\ref{Fig:5}.  

From Fig.~\ref{Fig:5} , we see that the activation energy of spins is given by 
the formation energy $E_{1}$, while that of charges by $E_{1}+E_{5}$
for their transport for a macroscopic scale, which is to be compared with the experimental result
of pressure dependences of Ref.~1. 
In this context, we note that $\alpha=\gamma_c/\gamma_\Delta = (U-2V)/(U-4V)$ 
will increase as the pressure increases, since the inter-site Coulomb interaction 
$V$ is more sensitive to pressure than on-site $U$.\cite{Nagaosa}
We see that the theoretical results in Fig.~\ref{Fig:5} and experiment go together, 
keeping in mind the following fact.
The conductivity, $\sigma$, due to the soliton excitation will be given by  
\begin{equation}
\sigma = A_c e^{-E_c/T} = e^{\ln A_c -E_c/T} = e^{-\varepsilon_c},
\end{equation}
where $A_c$ is a pre-exponential factor. 
Therefore, $T\varepsilon_c$ should be
\begin{equation}
T \varepsilon_c = E_c - T\ln A_c.
\end{equation}
The term $-T\ln A_c$ will affect the absolute value of $T\varepsilon_c$, 
but its pressure dependence will be weak because of logarithmic 
dependence on $A_c$, and then we expect that the pressure dependence of
$T\varepsilon_c$ reflects that of $E_c$. 
Similar considerations will apply to $E_s$ as well.

\bigskip\noindent
{\bf Acknowledgment}

Authors thank K.\ Kanoda for stimulating and detailed 
discussions on his experimental findings reported in Ref.~1.

\end{document}